# Effects of substrate temperature on the unusual non-Fermi liquid metal to insulator transition in perovskite SrIrO$_3$ thin films

Abhijit Biswas and Yoon Hee Jeong[a]

*Department of Physics, Pohang University of Science and Technology, Pohang, 790-784, S. Korea*

[a]Author to whom correspondence should be addressed: *yhj@postech.ac.kr*

**Abstract**

Electronic transport has been investigated for strong spin-orbit coupled perovskite SrIrO$_3$ thin films grown at various substrate temperatures. The electronic transport of the SrIrO$_3$ films is found to be very sensitive to the growth parameters; in particular, the film can either be a metal or an insulator depending upon the substrate growth temperature. While all the metallic films show unusual sublinear temperature dependent non-Fermi liquid behaviors in resistivity, the insulating film grown at a higher temperature stands out for its inhomogeneous Ir distribution, as analyzed by secondary ion mass spectrometry. This observation demonstrates that the inhomogeneous distribution of cations can be one of the fundamental factors in affecting the electronic transport in heavy element based oxide films and heterostructures.

(Some figures may appear in colour only in the online journal)

# 1. Introduction

Ruddlesden-Popper phases of strontium iridates $Sr_{n+1}Ir_nO_{3n+1}$ ($n$ = 1, 2, and ∞), have been a subject of current prolific investigations due to rich phase diagrams involving dimensionality controlled metal insulator transitions (MIT) arising from the interplay between strong spin-orbit coupling (SOC) and local Coulomb interactions [1]. In this dimensionality controlled MIT scenario, layered $Sr_2IrO_4$ ($n$ = 1) was illustrated to be a $J_{eff}$ = 1/2 antiferromagnetic Mott insulator [2]. $Sr_3Ir_2O_7$ ($n$ = 2), on the other hand, turns out to be barely an insulator, displaying a ferromagnetic state with a Curie temperature of 285 K [3]. With an increase in the number of $IrO_2$ planes, that is, with the increasing number of the $IrO_6$ octahedra layers ($n$), the bandwidth of Ir $5d$ states becomes broader. As a result, perovskite $SrIrO_3$ ($n$ = ∞) becomes a correlated bad metal and the mean free path becomes comparable to the inter-atomic distance, and thus $SrIrO_3$ is presumably close to a metal-insulator phase boundary [1].

Structurally, $SrIrO_3$ has the crystal structure of hexagonal $BaTiO_3$ at room temperature and atmospheric pressure; the perovskite structure is formed only at high pressure and temperature [4]. Studies on hexagonal $SrIrO_3$ single crystals have revealed that it exhibits non-Fermi liquid behaviors particularly in resistivity [5]. Till date perovskite $SrIrO_3$ single crystals, however, are not available as it needs high pressure synthesis techniques. In order to obtain perovskite $SrIrO_3$ and also to induce new physical phenomena different from bulk properties, one might take advantages of thin film growth technology. With thin films, not only metastable phases can be stabilized on preferred substrates but also key parameters such as band width, SOC, and correlation of the system can be adjusted. Recently several efforts were reported from thin films point of view in synthesizing perovskite $SrIrO_3$ and many new surprising physical phenomena have been observed [6-10].

One of the major advantages for thin films (not only for $SrIrO_3$ but also in general perspectives) is that their electronic properties can not only be tuned by varying underlying substrates but also be tuned by controlling the growth parameters for a given substrate. Investigating the electronic properties of thin films by varying the substrate growth temperature ($T_G$) while keeping the other growth parameters unchanged would be an important avenue to explore since the quality of thin films strongly depends on the substrate temperature during growth [11]. There are several reports about the growth temperature dependent electronic structure in thin films of $SrRuO_3$ with the analogous structure; for the films grown at various substrate temperatures, changes in electronic transports were observed [12-14]. In some cases, the structure of films may depend on growth conditions as shown for $PrVO_3$ films where the structure changes from orthorhombic to monoclinic [15]. Changes in the supply of external gas pressure (here oxygen) might also be a parameter to control, but since perovskite $SrIrO_3$ can only be stabilized at high oxygen pressures, keeping the oxygen partial pressure high and constant is an automatic choice. Also from applications point of view substrate growth temperature can influence the gas sensitivity and optical properties of thin films [16].

It is natural to expect that perovskite $SrIrO_3$ with SOC, correlation, and dimensionality operating together would be susceptible to external parameter changes and perhaps an appropriate perturbation would be able to push the system to an emerging new phase. Here, we show that increasing the substrate growth temperature pushes the system from an unusual non-Fermi liquid metallic state with sublinear temperature dependent resistivity to an insulating state. Moreover, it is further shown that this insulating phase is accomplished by the unique thin film feature that at higher growth temperatures, the cationic elemental distribution in the films becomes inhomogeneous and disordered, which in turn brings about a MIT.

## 2. Experimental details

We have grown perovskite $SrIrO_3$ thin films on best lattice matched substrate $GdScO_3$ (110) (the pseudocubic ($a_c$) lattice parameters of bulk $SrIrO_3$ and $GdScO_3$ were found to be the same ∼3.96 Å) while varying the substrate growth temperature from 500° C to 700° C with all the other parameters kept constant. $SrIrO_3$ thin film of typical thickness of ∼35 nm were fabricated by pulsed laser deposition (KrF laser with λ = 248 nm) from a polycrystalline target with stoichiometric 113 composition of Sr, Ir, and O. The target was made by mixing the stoichiometric amount of raw powders ($SrCO_3$ and $IrO_2$), pelletizing, and sintering at high temperature and ambient pressure. It may be noted here that the target is not in a single phase of perovskite $SrIrO_3$ because perovskite $SrIrO_3$ is not obtained with ambient pressure synthesis and would require high pressure treatments. During growth the laser was operated at frequency 4 Hz, and the oxygen partial pressure in the growth chamber were 20 mTorr. All the films were post annealed at the same oxygen partial pressure for 30 min to compensate for any oxygen deficiency. To check the crystalline quality, X-ray diffraction (XRD) measurements were performed by the Empyrean XRD System from PANalytical. Atomic force microscopy (AFM) was used for surface topography; it was done with a XE-100 Advanced Scanning Probe Microscope at room temperature. Electrical transport measurements were performed using the four-probe van der Pauw geometry. Ion homogeneity of the films was characterized as a function of depth by secondary ions mass spectrometry (SIMS) with a primary beam source of $O_2^+$ with impact energy of 7.5 keV.

## 3. Results and discussion

To start with, X-ray $\theta$–$2\theta$ scan for the SrIrO$_3$ films synthesized as described above show the crystalline (001)$_C$ peak without any impurity or additional peaks; for clarity, only the low angle data have been shown in Fig. 1. As can be seen in the figure, the film grown at 550° C displays nice thickness fringes with the substrate and film peaks located closely whereas these thickness fringes disappear and the peaks get broadened as the growth temperature deviates from 550° C. The peak broadening is quite remarkable and may indicate that the atoms in a unit cell may not be in exact order which accounts for the structure factor in the scattered intensity. The inhomogeneous distribution of the atoms in the unit cells would cause the Bragg peak to merge with the fringe peaks and make the observed peak appear to be broader for high temperature growth films. It is to be recalled that the diffraction peaks of X-ray scattering (Bragg peak as well as fringe peaks in the films) are the results of long range ordering only. Any contributions to the structure factor from disorder such as an inhomogeneous distribution of elements, oxygen vacancies, or a bilayer type structure with coherent and incoherent layers (coherent part corresponds to the Bragg peak only) would add up to the Bragg peak profile [17]. Presumably we may state that the amount of disorder in the SrIrO$_3$ films changes with the growth temperature and this change then leads to the distinct features in the XRD profile in Fig. 1. Our recent report has shown that the disorder in the form of Ir inhomogeneity would be one of the major sources of disorder in SrIrO$_3$ films [6].

To characterize the surface of the films, surface morphologies were obtained by atomic force microscopy (AFM) and the images are shown in Fig. 2. As shown in the Fig. 2(b), the surface is flat without any particles on it for the films grown at 550° C. With either decreasing or increasing the growth temperature, particles are found on the surface with a variation in height, 3-5 nm for films at 600° C as shown in Fig. 2(c) and 10-15 nm for films at 500° C, 650° C, and 700° C as shown in Fig. 2(a), 2(d) and 2(e). For more quantitative description, an exemplary line profile is shown in Fig. 2(f). To identify these particles or islands, it is reminded that strontium based particles (SrO) need a high temeprature (≥1000° C) for formation at the oxides surfaces and the growth tempratures used are too low for SrO to form [18]. Thus, the particles seem to be iridium based ones, possibly $IrO_2$. Also it is noted that the root mean square (rms) roughness of the films increases with the increase or decrease of the growth temperature from 550° C as indicated in the inset of each topography.

Now having analyzed the crystalline qualities, let us turn to the electrical transport measurements. Figure 3 shows the transport behavior of $SrIrO_3$ films on $GdScO_3$ (110) grown at various temperatures. As shown in the figure, the resistivity of the films increases from a lower value to a higher one as we increase the growth temperature from 550° C. In other words, the resistivity at room temperature is minimal for film grown at 550° C, and this is consistent with the fact that the degree of disorder is much more prominent in other films as characterized by x-ray analysis and surface morphology measurements. When the growth temperature is further increased to 700° C, the film reaches an insulating state. It is mentioned here that the film grown at a lower temperature 500° C shows a much higher resistivity value than the best one does (550° C) and also has more rough surface morphology as shown earlier. Currently the microscopic origin behind this is unknown, but it may be argued that in this case, the poorer crystalline quality is due to lower velocity of crystallization since cooling rate of the condensate becomes high at lower substrate temperature [11].

Quantitatively speaking, the resistivity of the best crystalline film grown at 550° C was found to be $\rho$ = 1.4 mΩ·cm at room temperature (inset of Fig. 3). This value is smaller than the bulk value 4 mΩ·cm of polycrystalline $SrIrO_3$ at room temperature; this fact can be understood by noting that the measured resistivity in bulk polycrystals is influenced by grain boundaries and porosity [19]. While the films grown at other temperatures also show metallic characteristics, an upturn in resistivity below 15 K was observed for the films grown above 650° C. This gives us an indication that disorder related Anderson localization plays a role in determining the transport properties of the SrIrO$_3$ films. We may emphasize here that the resistivity variation as a function of temperature for a given film is rather small ($\rho_{300K}/\rho_{10K}$ ~ 1.5) and thus the SrIrO$_3$ films would be regarded as a bad metal [6,20]. When the growth temperature is further increased to 700° C, the film shows a fully insulating behavior in electrical resistivity. At first glimpse, it would be seen that at a higher growth temperature the amount of disorder incorporated in the film is high enough to induce Anderson localization which in turn leads the system to an insulating state. Thus, it is concluded that growth temperature dependent MIT is observed in perovskite SrIrO$_3$ thin films.

For a deeper understanding of MIT, let us take a more detailed look at the transport properties of the metallic films. The most surprising feature about the metallic transport behaviors of the perovskite SrIrO$_3$ films is the *sublinear in temperature* dependence of the electrical resistivity over a wide range of temperature (300 K ~ 10 K) as displayed in Fig. 4(a)-(d) [6]. The resistivity of the best crystalline film grown at 550° C varies as $\rho \propto T^{0.80}$, as displayed in Fig. 4(b). From fitting the data with the formula $\rho = \rho_0 + A \cdot T^{0.80}$, we obtain the residual resistivity $\rho_0 = 0.969$ mΩ·cm and $A = 0.00436$ in proper units with resistivity in mΩ·cm and temperature in K. The same sublinearity feature persists even for the films grown at 600° C as shown in Fig. 4(c). Fig. 4(a) and 4(d) also illustrate that a similar power law variation in resistivity ($\rho \propto T^{0.75}$) holds for the films grown at 500° C and 650° C despite the fact that the degree of disorder is different in these films.

Recently it was shown that changes in the amount of disorder in the presence of spin fluctuations in some metals near the quantum critical point can alter the power law in resistivity [21]. Perhaps disorder in the metallic SrIrO$_3$ films is also playing a key role in changing the power law in resistivity [6]. Also the sublinear temperature dependence over a wide range of temperature observed in the present system is extremely remarkable because it cannot be explained within the conventional quantum critical scenario of the Hertz-Moriya-Millis framework [22]. It is to be noted that we were not able to find any evidence of magnetic long range ordering in the measured temperature range for these films. Despite lack of magnetic ordering in the system, we conjecture that localized electrons may give rise to local magnetic moments without collective magnetic fluctuations since they exist near the boundary of a metal-insulator transition. Such localized moments or small magnetic clusters can influence the electronic transport significantly, which is expected to be described within a dynamical mean-field theory (DMFT) framework [23]. Moreover for the insulating film grown at 700° C, the resistivity below approximately 15 K can be fitted well within the conventional three dimensional variable range hopping (3$d$-VRH) model (ln $\sigma \propto 1/T^{1/4}$) as shown in Fig. 4(e). The fitting of the data with the VRH model indicates that disorder plays a significant role for the films grown at the high temperature.

The unusual sublinear non-Fermi liquid behavior and the evolution of MIT with a resistivity upturn in between observed in the present system seem to be connected with disorder. In fact, this connection would be a strong possibility as previous studies have shown that if strength of disorder ($D$) crosses a certain limit ($D > D_{NFL}$) in a correlated system, then it enters into a "*Griffiths phase*" displaying metallic non-Fermi liquid behaviors and even with stronger disorder above some critical value $D_C$ ($D > D_C > D_{NFL}$) the system undergoes a MIT [24]. It would be a challenging task to construct a proper theory which accounts for the sublinear variation in resistivity in the presence of disorder. A detailed discussion about the "*Griffiths phase*" type scenario (inhomogeneous formation of Mott insulating islands and Fermi liquid metal islands) can be found elsewhere [6]. Recent theoretical calculation has also shown that if both Hubbard interaction and the strength of spin orbit couplings are tuned, various ground states can be achieved for bulk perovskite $SrIrO_3$ [25]. It is then highly desirable to pursue such calculations by considering disorder (*D*) as another varying parameter in the presence of interaction (*U*) and spin orbit coupling (*SOC*).

To get a clue on the source(s) of possible disorder responsible for this growth temperature tuned MIT phenomenon, we analyzed each film's homogeneity by secondary ions mass spectrometry (SIMS). Previously, we already observed a MIT due to Ir inhomogenity in highly strained films [6]. SIMS is a powerful tool to obtain a depth wise homogeneity of a specific element. As shown in the Fig. 5, the films show more or less homogeneous Ir distribution except for the film grown at high temperature of 700° C. Depth profile analysis of the other elements (Sr and O) also reveals that they are homogeneous for all the films. Thus, we may conclude from the SIMS results that the film grown at 700° C possesses an inhomogeneous distribution of heavy element Ir causing a large amount of disorder, consistent with the insulating behavior in resistivity. Inhomogeneity in the Ir distribution would mean that Ir atoms are displaced from their periodic positions and this would require some excess activation energy and/or vacancies through which the ions can migrate. High substrate growth temperature can provide this energy necessary to activate such migration [11]. Thus, with the increase in substrate growth temperature, the rate of Ir migration and the consequent amount of disorder increase. It is worth mentioning that the inhomogeneity in the Ir distribution may also annihilate the charge balance and make the structure unstable. In addition to influencing the electronic properties such as resistivity, the presence of the disorder would also mean the existence of a certain density of defects in the system and thus influence the crystalline quality of the films as seen from Fig. 1, which in turn causes a broadening of the XRD profiles in the films. The XRD peak broadening and an increase of particle sizes on the surface are consistent with the increasing amount of defects accompanying the structural disorder. It may be mentioned that the Ir inhomogeneity along the deposition direction would probably exist in other films grown at lower temperatures as well (as resistivity increases indicate) although it is not seen at the resolution level of SIMS measurements.

## 4. Conclusions

In summary, a MIT from a non-Fermi liquid state with resistivity sublinear in temperature to an insulator state was observed in perovskite $SrIrO_3$ thin films as the substrate temperature for film growth changes. The peculiar sublinear temperature dependent resistivity may originate from local magnetic fluctuations which would need further investigations via some local probes and we are currently working in this direction. The observed MIT may be attributed to the fact that the cationic elemental distribution becomes inhomogeneous creating scattering centers and thus the inhomogeneous distribution would localize electrons and induce an insulating feature in resistivity of films grown at higher temperatures. Atomic level investigations on how the cationic distribution becomes inhomogeneous as well as systematic cationic compositional analysis for a series of the films with increasing growth temperature would be a topic for future research. Also it would be interesting to study experimentally as well as theoretically whether this growth temperature induced inhomogeneity appears in general for other $5d$ heavy element (Os, Re etc.) based thin films and heterostructures. Specifically for the $SrIrO_3$ case, it would be of interest to check whether high growth temperature is always preferred for the perfect crystalline thin film growth on various substrates. The present observations of the transport properties of $SrIrO_3$ films clearly illustrate that growth temperature induced elemental inhomogeneity can play a crucial role and become one of the major factors affecting the electronic transport in heavy element based oxide films and heterostructures. In closing we wish to stress that our current understanding of the non-Fermi liquid physics of SrIrO$_3$ or other systems involving 5$d$ heavy elements is far from being complete and hope the present work catalyze subsequent studies.

**Acknowledgments**

We would like to thank Prof. J. H. Park and Prof. J. S. Kim for providing X-ray facility to us, NINT Korea for SIMS measurements and Y. W. Lee and S. W. Kim for technical helps. This work was supported by the National Research Foundation (Grant No. 2011-0009231) and the Center for Topological Matter at POSTECH (Grant No. 2011-0030786).

**Figure Captions:**

**Figure 1.** (Color Online) X-ray $\theta$–$2\theta$ scan results of SrIrO$_3$ films grown on lattice matched GdScO$_3$ (110) substrates at various growth temperatures. The (001) pseudo-cubic film peaks and the substrate (110) peaks are seen.

**Figure 2.** (Color online) (a)-(e) Atomic force microscopy images of the surface of the films grown at various growth temperatures. Upper right inset of each figure shows the growth temperature (T$_G$) and lower right inset the root mean square (rms) roughness of the films. (f) The line profile obtained across the red line of figure (e) shows the particles size. (Scan size: $5 \times 5$ μm$^2$).

**Figure 3.** (Color online) Temperature dependent resistivity of the films grown at various growth temperatures. While the films grown at lower temperatures are metallic, an insulating behavior is observed for the film grown at 700° C. Room temperature resistivity is shown as a function of T$_G$ in the inset.

**Figure 4.** (Color online) (a)-(d) Sublinear in temperature dependent resistivity of the films grown at various temperatures. While the films grown at 550° C and 600° C shows T$^{0.80}$ dependence, T$^{0.75}$ variation in resistivity is observed for the films grown at 500° C and at 650° C. (e) The resistivity of the insulating film grown at 700° C follows the three dimensional variable range hopping model.

**Figure 5.** (Color online) Depth profile of the Ir distribution of the films grown at various temperatures as analyzed by secondary ion mass spectrometry. Inhomogeneous distribution of Ir is found for the film grown at 700° C. The vertical dotted line indicates the film/substrate interface.

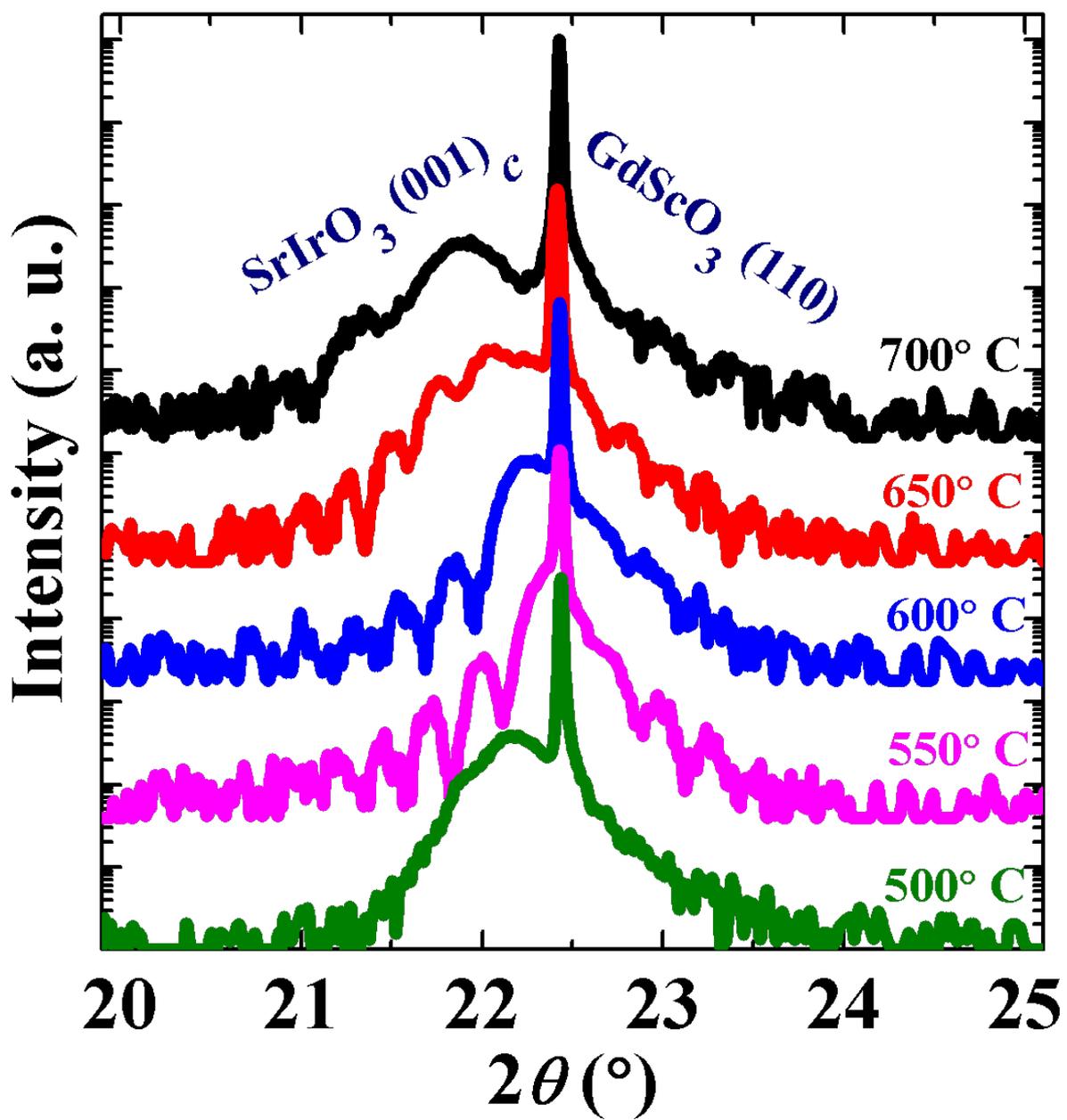

**(Figure 1)**

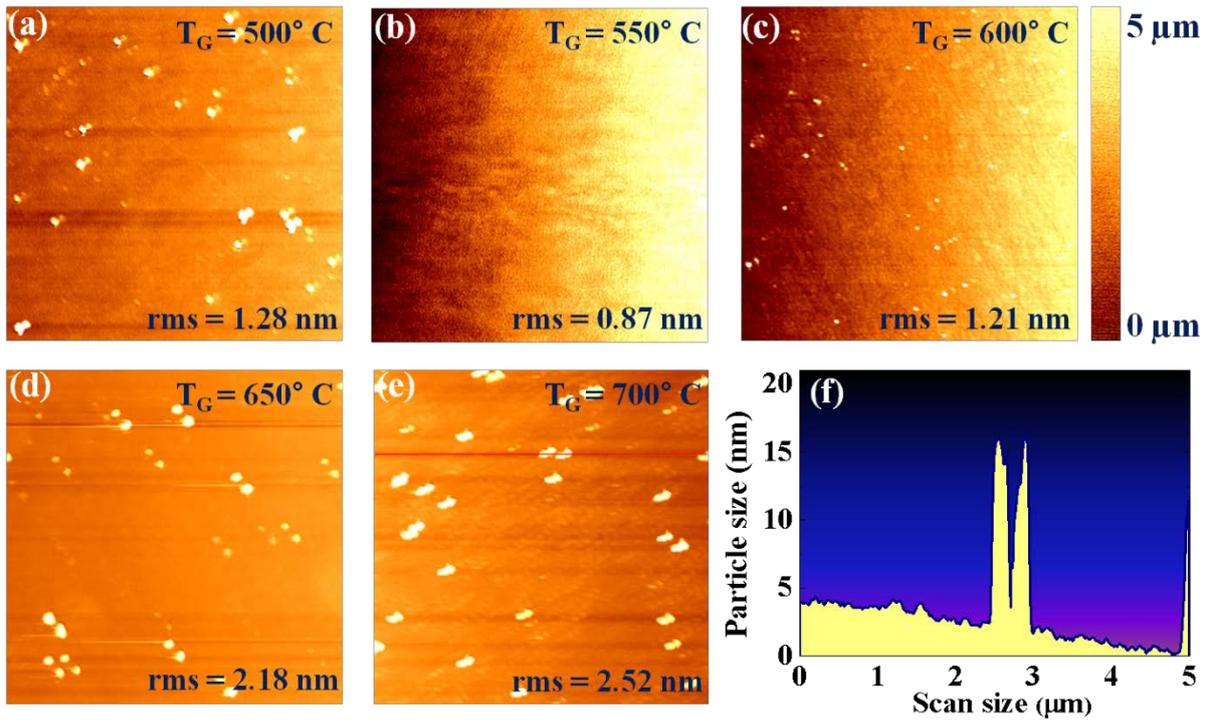

**(Figure 2)**

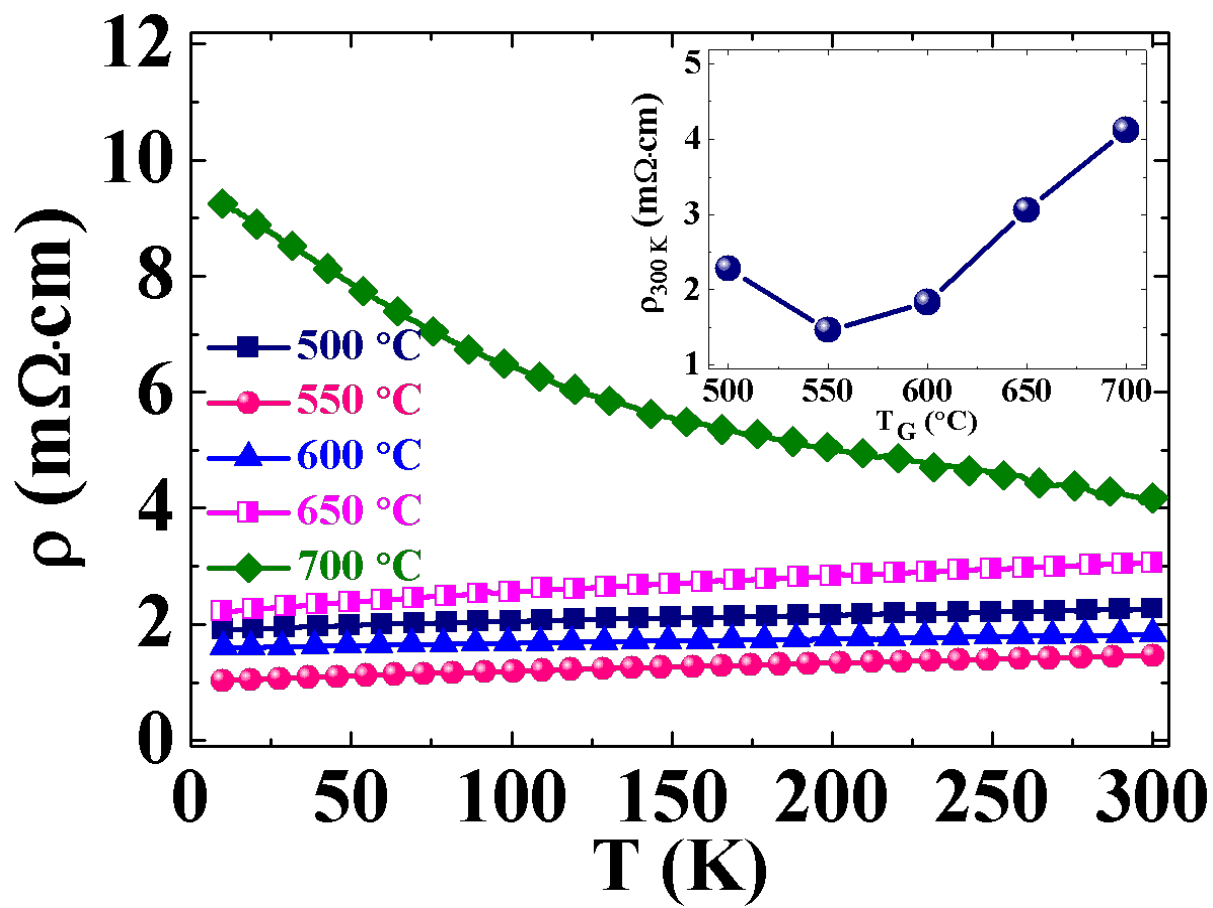

(Figure 3)

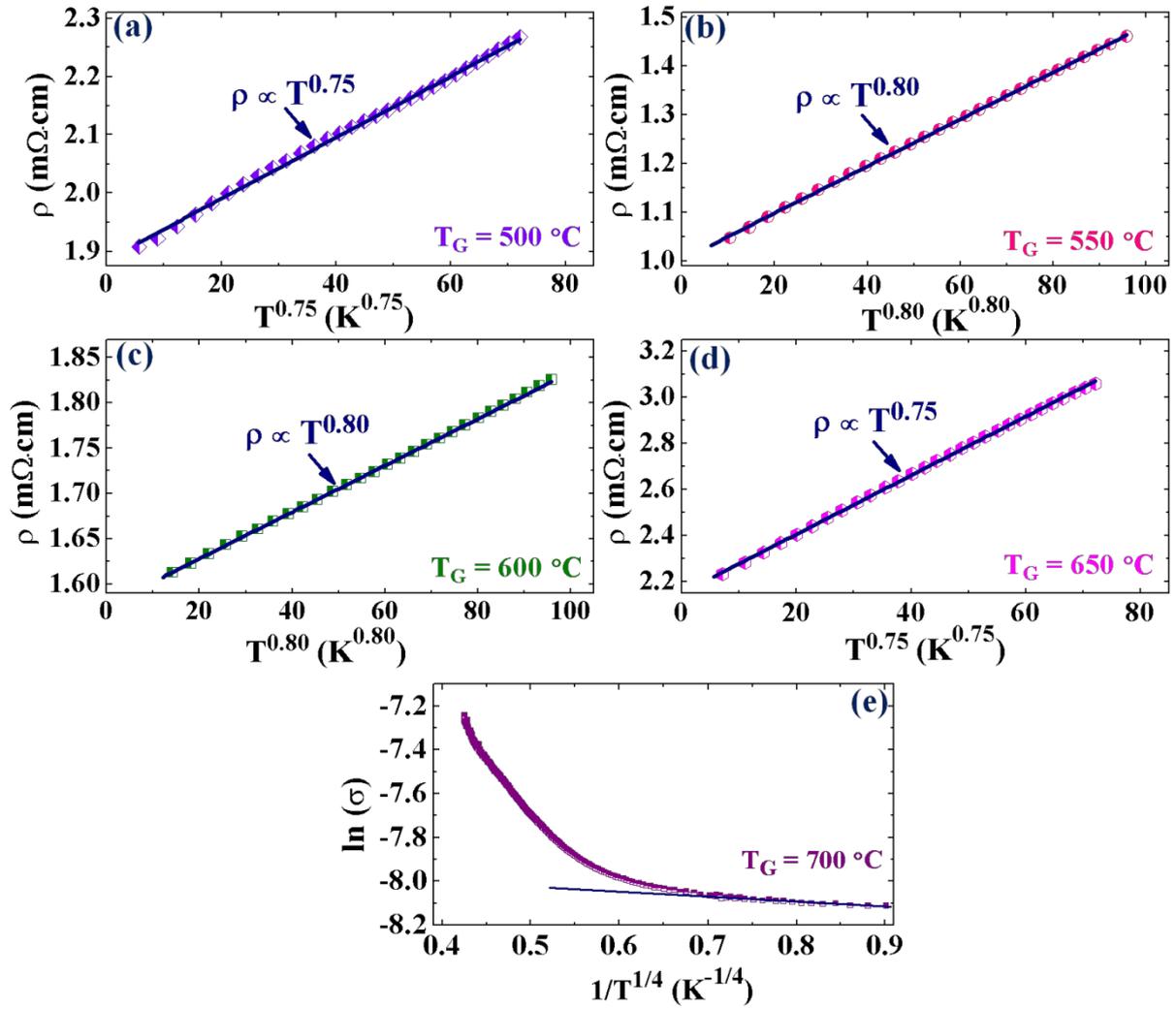

**(Figure 4)**

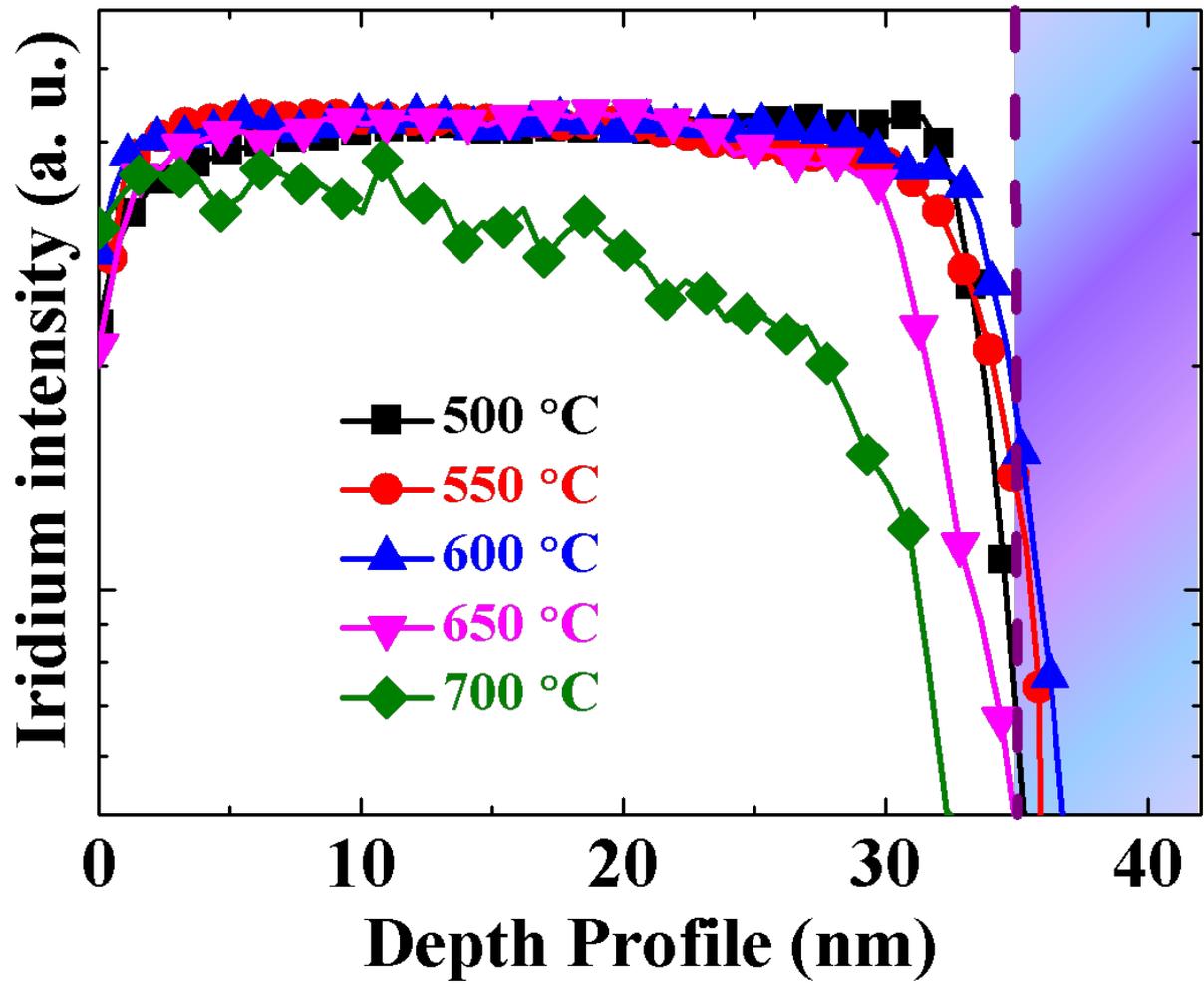

(Figure 5)